\documentclass[aps,prb,longbibliography,floatfix,reprint]{revtex4-1}
\usepackage{amsmath}
\usepackage{graphicx}
\usepackage{amsfonts}
\usepackage{color}
\usepackage{placeins}

\begin{document}

\title{Analysis of nonlocal phonon thermal conductivity simulations showing the ballistic to diffusive crossover}

\author{ Philip B. Allen }
\email{philip.allen@stonybrook.edu}
\affiliation{ Department of Physics and Astronomy,
              Stony Brook University, 
              Stony Brook, New York 11794-3800, USA }

\date{\today}

\begin{abstract}

Simulations ({\it e.g.} Zhou {\it et al.}, Phys. Rev. B {\bf  79}, 115201 (2009)) show nonlocal effects of the ballistic/diffusive
crossover.  The local temperature has nonlinear spatial variation not contained in the local Fourier law 
$\vec{j}(\vec{r})=-\kappa\vec{\nabla}T(\vec{r})$.
The heat current $\vec{j}(\vec{r})$ depends not just on the local temperature
gradient $\vec{\nabla}T(\vec{r})$, but also on temperatures at points $\vec{r}^{ \ \prime}$ within phonon mean
free paths, which can be micrometers long.  This paper uses the Peierls-Boltzmann transport theory
in non-local form to analyze the spatial variation $\Delta T(\vec{r})$.  The relaxation-time approximation (RTA) is used
because full solution is very challenging.  Improved methods of extrapolation to obtain the
bulk thermal conductivity $\kappa$ are proposed.  Callaway invented an approximate method of correcting RTA for the 
$\vec{q}$ (phonon wavevector or crystal momentum) conservation of N (normal as opposed to Umklapp) 
anharmonic collisions  This method is generalized to the non-local case where $\kappa(\vec{k})$ depends on wavevector
of the current $\vec{j}(\vec{k})$ and temperature gradient $i\vec{k}\Delta T(\vec{k})$.  
\end{abstract}

\maketitle

\section{Introduction}

\begin{figure}
\includegraphics[angle=0,width=0.49\textwidth]{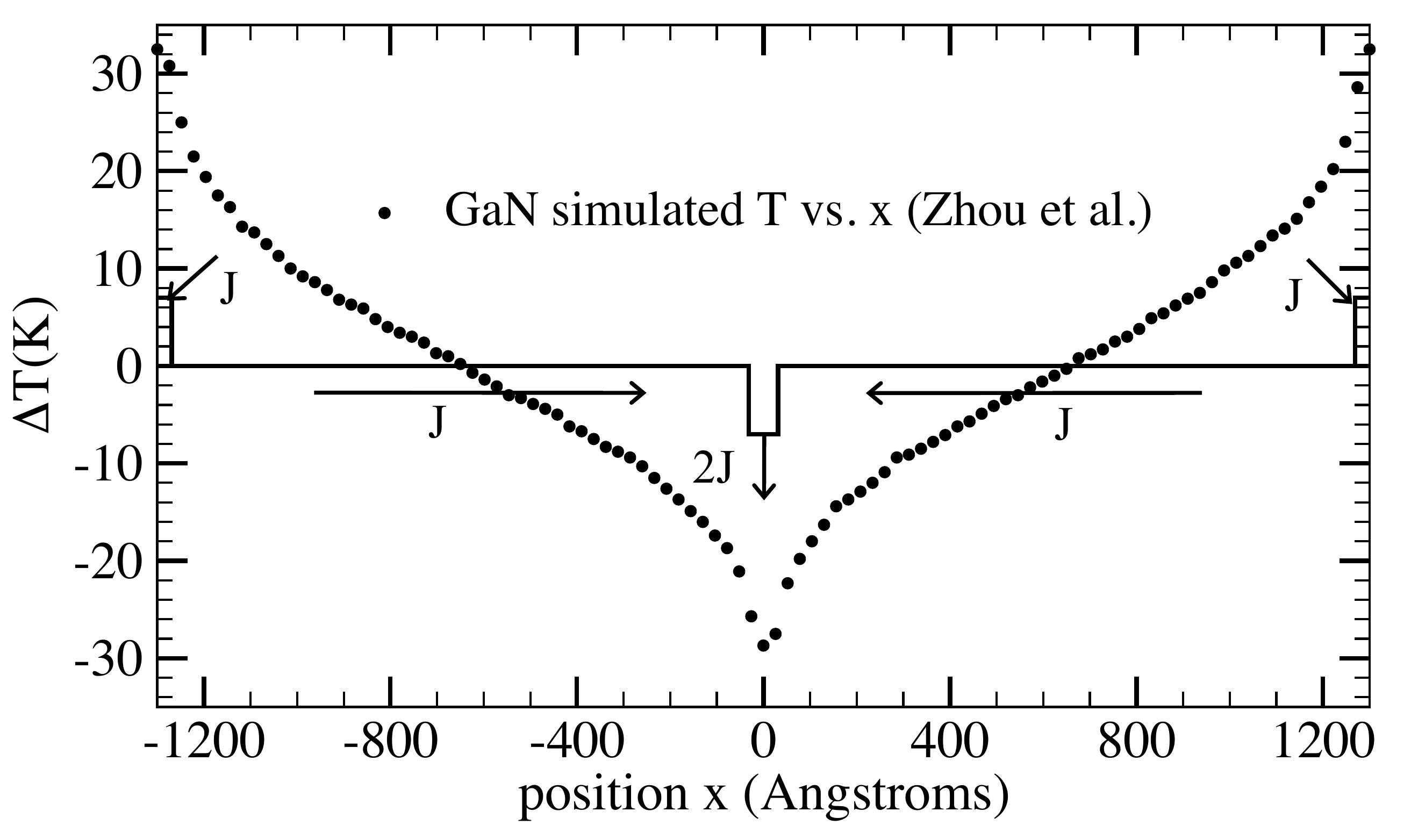}
\caption{\label{fig:Tofx} Heat flow simulation by Zhou {\it et al.} \cite{Zhou2009} for
wurtzite-structure GaN.  A segment of length $L = 500c = 2600\AA$ (along the $c$-axis) and cross
section $A = 15\sqrt{3}a^2=264\AA^2$, containing 60,000 atoms, was periodically repeated in all directions.
The average temperature was $\bar{T}=$301.2K.
Heat $Pd=0.003$eV/ps$\AA^2$ (volumetric heating rate $7.7\times 10^{18}$W/m$^3$) was added at segments of
width $d=12c$ at $-L/2$ (and equivalently at $+L/2$) and extracted at $L=0$.  
These regions are shown by the solid black lines.  In the regions of length $L/2-d$ between heat
insertion and removal, a constant heat current $j=0.0015$ eV/ps$\AA^2=2.4\times 10^{10}$W/m$^2$ flows. 
The temperature was averaged in 100 discrete segments of width $5c$.  
The gradient $dT/dx=\pm 0.0265$K$/\AA$ computed at midpoints
$\pm L/4$ corresponds to $\kappa(L)=-j/(dT/dx)=$90.4W/mK.  This number will increase by a significant
amount (perhaps a factor more than 2) for a very long simulation cell.  An ``effective'' thermal conductivity
can be defined by using the mean temperatures $T_H$ and $T_C$ of the regions of heating and cooling.
Then $(T_H-T_C)/(L/2)$ is a mean temperature gradient, giving $\kappa_{\rm eff}=55$W/mK.}
\end{figure}

Figure \ref{fig:Tofx} shows a molecular dynamics (MD) simulation for insulating wurtzite-structure 
GaN.  It shows the spatial change of temperature $\Delta T(x)$
driven by steady heat input $P(x)$ in regions near $x=nL \pm L/2$ and equal steady heat removal near $x=nL$.  This MD
study by Zhou {\it et al.} \cite{Zhou2009} illustrates nicely the nonlocal relation between temperature
gradient $\nabla_x T$ and heat current $j(x)$.  
\par
\par
Because of the care and accuracy of the simulation, and 
also because the system studied was periodic (with period $L$ which enables Fourier space analysis
with discrete wavevectors $2\pi m/L$), it is nicely suited for deeper analysis.  The current paper argues that the spatial
variation of $\Delta T(x)$ is the property most interesting for study, not the ``effective conductivity'' or
similar constructs that may be more easily measurable.  In a separate paper \cite{Allen2018} the concept
of ``thermal susceptibility'' ($\Theta$ where 
$\Delta T(\vec{r})=\int d\vec{r}^{ \ \prime}\Theta(\vec{r}-\vec{r}^{ \ \prime})P(\vec{r}^{ \ \prime})$) is introduced.
It has an inverse relation to thermal conductivity, much as the charge susceptibility has an inverse
relation to electrical conductivity.

In small insulators, with size $L$ similar to the long mean free paths $\ell$ of small $|\vec{q}|$ (long wavelength) 
acoustic phonons, heat transport deviates from the local Fourier law $\vec{j}(\vec{r})=-\kappa\vec{\nabla}T(\vec{r})$.
This is topic has attracted attention for more than 25 years \cite{Cahill2014,Majumdar1993,Joshi1993,
Chen1998,Chen2001}.  The terminology ``ballistic/diffusive crossover" is common.  
Diffusive heat propagation gives the local Fourier law, but
ballistic heat propagation requires (in linear approximation) a nonlocal kernel 
$\vec{j}(\vec{r})=-\int d\vec{r}^{ \ \prime} \kappa(\vec{r},\vec{r}^{ \ \prime})\vec{\nabla}T(\vec{r}^{ \ \prime})$. 
The range of the kernel is $|\vec{r}-\vec{r}^{ \ \prime}|\sim \ell$.  Interesting (and technologically important) non-local effects
happen if $\vec{r}$ is within a distance $\ell$ of a heat source or sample boundary.

Advances in measurement \cite{Gomes2015} include 
coherent x-ray thermal probing of strip-line arrays \cite{Siemens2010,Hoogeboom-Pot2015,Zeng2015},
transient thermal gratings \cite{Maznev2011,Johnson2013,HuaA2014,Huberman2017}, and
time-domain thermal reflectance \cite{Minnich2011,Ding2014,Wilson2014}.
Theory has become increasingly powerful \cite{HuaB2014,Vermeersch2016,Vermeersch2018} and
evolves togther with experiment \cite{Minnich2012,Collins2013,Minnich2015}.

My tool for analysis of the ``data'' of Fig. 1 is the Peierls-Boltzmann equation (PBE\cite{Peierls1929,Ziman1960}).
The PBE must treat explicitly the sources and sinks of heat, since they are at distances $\sim L/4$ from the
source, and this is not larger than $\ell_Q$ for many important phonon modes $Q$.  My analysis 
benefits from recent improvements, which include an explicit heat source term in the PBE \cite{HuaB2014,VermeerschI2015},
and a Fourier-transformed ($\vec{k}$-space) version of the PBE \cite{HuaB2014,Allen2014}.  
The PBE is not as microscopic as
an MD simulation, which can approximate exact atom-level motions (treated classically).  Boltzmann
theory, on the other hand, uses phonon quasiparticles.  The particle (rather than wave) picture is used, 
and requires wave-packets.  The spatial resolution of Boltzmann theory is thus limited
by the size of the wave-packet, {\it i.e.} not shorter than a phonon wavelength.  Crystalline matter is
spatially inhomogeneous at the atomic level, but spatially homogeneous at length scales greater
than lattice constants.  Thus Boltzmann analysis of crystals gives a non-local thermal conductivity
$\kappa(\vec{r},\vec{r}^{ \ \prime})$ invariant under the simultaneous translation $\vec{r}\rightarrow\vec{r}+\vec{s}, \ \
\vec{r}^{ \ \prime}\rightarrow\vec{r}^{ \ \prime}+\vec{s}$.  In atomic level theory, the translations $\vec{s}$  
are the lattice translation vectors.  However, the PBE is insensitive to this distance scale, so $\vec{s}$ can
be regarded as arbitrary and continuous.  Thus $\kappa$ can be written as $\kappa(\vec{r}-\vec{r}^{ \ \prime})$, 
and can be represented in $\vec{k}$-space as $\kappa(\vec{k})$, rather than as
$\kappa(\vec{k}+\vec{G},\vec{k}+\vec{G}^\prime)$.  On the other hand, the PBE treats phonons as
quantum objects, and is thus not limited to the high $T$ classical limit where MD simulation works.

In this paper, $Q$ will always refer to the quantum numbers $(\vec{q},s)$ of
a phonon, and $\vec{k}$ (or $k=|\vec{k}|$) will denote the reciprocal space coordinate
(or wavevector) of a field (like heat current, $\vec{j}$).   The same symbol
is used for functions (like $j(x)$ or $\kappa(x-x^\prime)$) in coordinate space and
in reciprocal space ({\it e.g.} $j(k)$ and $\kappa(k)$).  The three-dimensional Fourier relations are
defined as $j(\vec{r})=(1/N)\sum_{\vec{k}}j(\vec{k})\exp(i\vec{k}\cdot\vec{r})$
and $j(\vec{k})=(1/\Omega_{\rm cell})\int d\vec{r}j(\vec{r})\exp(-i\vec{k}\cdot\vec{r})$, with
$N$ the number of unit cells in the crystal and $\Omega_{\rm cell}$ the volume of the crystal primitive cell.
This paper concerns nanoscales in one direction, chosen as $x$ (as in Fig. \ref{fig:Tofx}).  The other directions
are macroscopic.  Therefore reciprocal space behavior involves $k_y=k_z=0$, and $k_x$ is abbreviated as $k$.

Fig. \ref{fig:Tofx} is a steady state nonequilibrium molecular dynamics (NEMD) simulation 
for GaN, with steady heat insertion and removal at a controlled rate,
in widely separated and narrow spatial regions.  Therefore the heat current $j(x)=j$ is steady, and known,
in the regions between heat insertion and removal.  Therefore the curvature of $T(x)$
evident in Fig. \ref{fig:Tofx} is a clear sign of a
nonlocal connection between $j$ and $T$.  The system had average $T\approx300$K; 
classical trajectories were computed from an empirical interatomic force law.
Local temperatures $T(x_i)=<KE>_i /3Nk_B$ were computed by time averaging the kinetic
energy $KE$ of all $N_i$ atoms in slabs (labeled $i$) containing $N_i=600$ atoms.  The aim
of the simulation was to make the total length $L$ large enough to achieve the diffusive
limit, so that $\kappa$ could be found by computing $\Delta T/\Delta x$ in central slabs.
However, GaN at 300K has many phonons with mean free paths exceeding the sample size $L$,
so the fully diffusive limit was not reached.  Extrapolation was attempted by the model
$\kappa(L)\approx\kappa(\infty)-\kappa^\prime/L$.  The current paper provides better extrapolation models.  
The extrapolated value was well below the experimental 230 W/mK
\cite{Slack2002,Jezowski2003}.  Interestingly, theory \cite{Lindsay2012} now shows that isotopically
pure GaN (as assumed in the simulation) should have $\kappa$(300K)$\approx$400 W/mK.  It is not
clear whether finite size or an inadequate model potential $V(R)$ is the main culprit
limiting the realism of the simulation.  However, for the purpose of this paper, material-specific realism
is irrelevant.  The model is useful for studying nonlocality, because of the care and accuracy of
the simulation, independent of possible problems with the potential.

The curvature of $T(x)$ seen in the figure was
regarded as a nuisance or an artifact of the finite size.  The alternative view advocated here
is that Fig. \ref{fig:Tofx} simulates an idealized experiment, not yet achievable. 
This ``experiment'' reveals details of nonlocality, and probes nicely the ballistic to diffusive crossover.

This simulation is simple to analyze for two reasons.  (1) Periodicity $f(x+L)=f(x)$ is maintained for
all fields.  (2) Regions of heat input and extraction are ``transparent'' to propagating phonon modes.  A
thermostat occasionally perturbs atom trajectories in discrete regions, but does not alter the
lattice periodicity that gives homogeneously propagating phonon modes.
The response function $\kappa(x-x^\prime)$ has periodicity $L$ in $x-x^\prime$,
and wavevectors in $\kappa(k)$ are quasi-discrete ($k=2\pi n/L$) and Bloch-periodic ($k\equiv k+2\pi/a$).
If instead a simulation had a hard wall or other disruption of homogeneity,
the Boltzmann equation would not separate when written in $k$-space, and
numerical solution to find $T(x)$ would be challenging.
Most actual nanoscale heat transport involves spatial inhomogeneity.
However, exceptions such as ``transient thermal grating'' (TTG) experiments \cite{Rogers1994}, and the
idealized experiments analyzed by Hua and Minnich \cite{Hua2018}, can be analyzed by the method used here.

Accurate solutions of the full PBE for bulk thermal 
conductivity have been available for several years \cite{Broido2007,Chernatynskiy2010,Li2014}.   
Several recent papers \cite{Koh2014,Miranda2015,HuaB2014,Hua2015,Maassen2015,Kaiser2017,Cepellotti2017a,Cepellotti2017}
find solutions of  the PBE containing nonlocal effects.

\begin{figure}
\includegraphics[angle=0,width=0.5\textwidth]{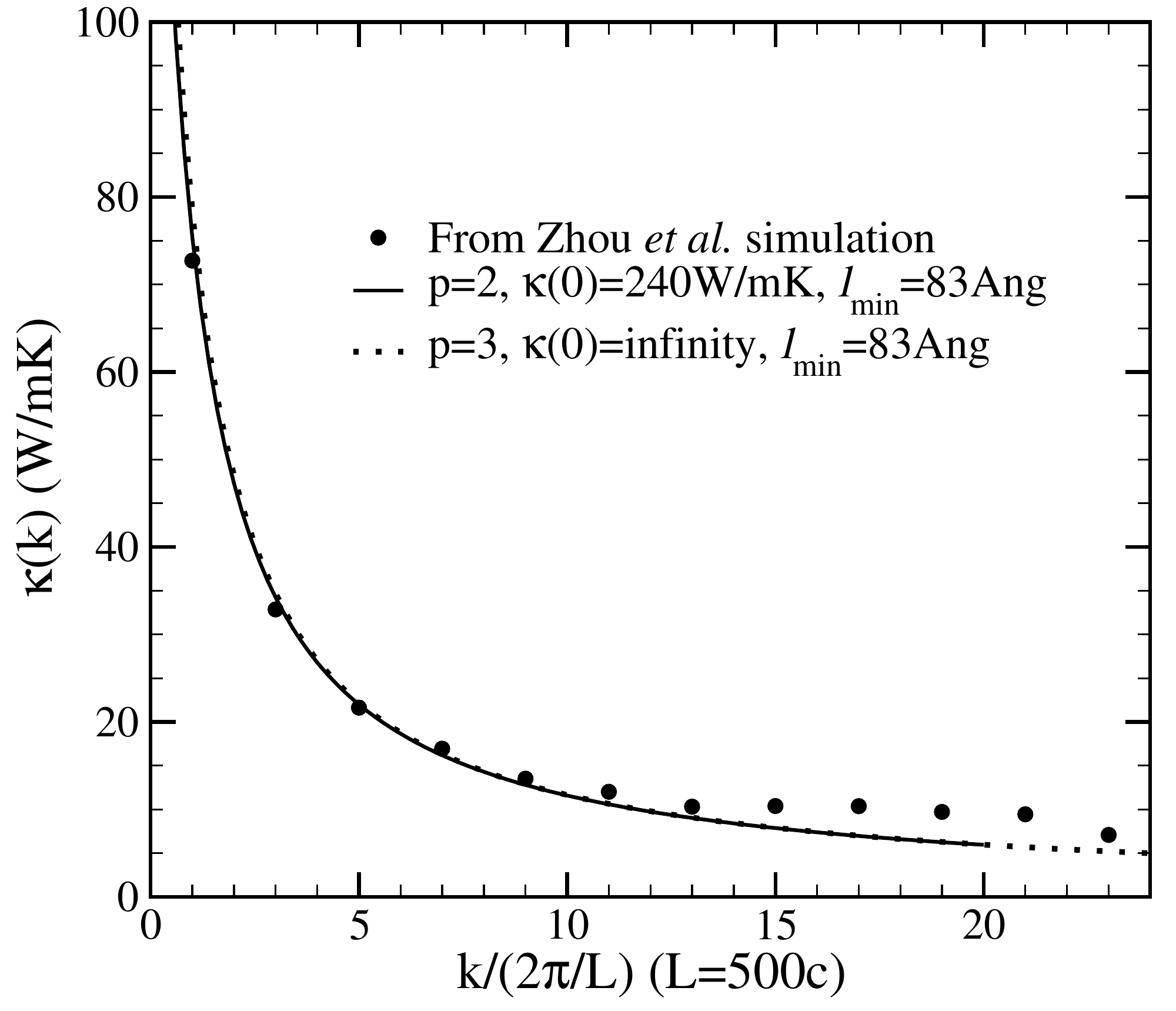}
\caption{\label{fig:calfit} The dots are nonlocal $\kappa(k)$ constructed for GaN at 300K,
by Fourier transforming the $T(x)$ results
shown in Fig. \ref{fig:Tofx}.  The curves are Debye-model RTA theoretical fits, discussed in Sec. \ref{sec:Debye}.
The same two adjustable parameters ($\kappa_0=$80 W/mK and $\ell_{\rm min}=83\AA$) are used in both curves. 
If the phonon scattering rate $1/\tau_Q\propto \omega^2$
({\it i.e.} $p=2$) is used, the bulk limit $\kappa(k\rightarrow 0)$ is $3\kappa_0=240\AA$.
If exponent $p=3$ is chosen, $\kappa(k)$ diverges as $|\log(k)|$ as $k\rightarrow 0$.}
\end{figure}

The outline of this paper is: Section \ref{sec:FT} discusses the discrete Fourier transform
used to convert Fig. \ref{fig:Tofx} to Fig. \ref{fig:calfit}.  Section \ref{sec:Boltz} discusses the nonlocal PBE,
and solves it using the relaxation-time approximation (RTA).  Section \ref{sec:Debye} gives numerical 
answers using a Debye model.  Section \ref{sec:kC} gives the nonlocal generalization of Callaway's approximation
for Normal (N) and Umklapp (U) collisions.  Section \ref{sec:extrap} shows how best to extrapolate simulation data
to the bulk limit.  In Appendix A, boundary-condition
influences on nanoscale non-locality are discussed.   Appendix B gives detailed analytic formulas for
various versions of the Debye model, and Appendix C gives details of the Callaway version
of the theory.

\par
\par

\section{Fourier transforms}
\label{sec:FT}

The information in Fig. \ref{fig:Tofx} is Fourier transformed following Ref. \onlinecite{Allen2014}. 
The resulting values of $\kappa(k)$ are shown in Fig. \ref{fig:calfit}.  Because the simulation
of Fig. \ref{fig:Tofx} has $M$=100 discrete segments,
the wavevector $k$ must have only $M=100$ possible values
$k_n=2\pi n/L$ for $-M/2+1 \le n \le M/2$, where $L/M=w=5c$ is the width of the separate segments
where $T$ is averaged.  But there are only $M/4$ independent real numbers in the computed $\Delta T(x)$, 
since $\Delta T(-x)$ converges to the same value as $\Delta T(x)$, and $\Delta T(L/4 + x)$
converges to the same value as $-\Delta T(L/4 - x)$ (barring small non-linear effects).
Therefore there are only $M/4=25$ real numbers in the Fourier representation, which can
be taken as the values of $\kappa(k_n)$ for positive odd integers $n$.   Only the smallest 12 $k_n$'s are shown
in Fig. \ref{fig:calfit}.  Higher $k_n$'s are increasingly noisy.  
Partly this is caused by noise in the original calculations, and partly by additional noise in the digitization (original
numerical information was not available.)

\section{Boltzmann $\kappa(k)$}
\label{sec:Boltz}

GaN is a good thermal conductor; its phonons have long mean free paths.
Thus it is a good ``phonon gas'' and should be accurately treated by the 
Boltzmann equation.  The fundamental object of Boltzmann theory is the distribution function $N_Q$, 
which gives the average occupation at $(\vec{r},t)$ in coordinate space, or $(\vec{k},\omega)$ in
reciprocal space, of phonon mode $Q$.  Its evolution is given by the equation,
\begin{equation}
\frac{\partial N_Q}{\partial t}=\left(\frac{dN_Q}{dt}\right)_{\rm drift}+\left(\frac{dN_Q}{dt}\right)_{\rm scatt}
+\left(\frac{dN_Q}{dt}\right)_{\rm ext}.
\label{eq:pbe}
\end{equation}
The result shown in Fig. \ref{fig:Tofx} has reached steady state in a time-independent thermal driving,
so $\partial N_Q/\partial t=0$.  The driving is one-dimensional, so the resulting current density is
\begin{equation}
j(x)=\frac{1}{\Omega_S}\sum_Q \hbar\omega_Q v_{Qx} N_Q(x) 
\label{eq:JJ}
\end{equation}
where $\vec{v}_Q$ is the group velocity of mode $Q$ and $\Omega_S=N\Omega_{\rm cell}$ is the sample volume.  
An identical equation applies to the reciprocal space relation between $j(k)$ and $N_Q(k)$.

The scattering term in a non-metal includes
defect scattering which couples $N_Q$ to $N_{Q^\prime}$, and
anharmonic scattering which couples $N_Q$ to ($N_{Q+Q^\prime}$, $N_{-Q^\prime}$).  
A local equilibrium Bose-Einstein
distribution $n_Q=[\exp(\hbar\omega_Q/k_B T(x))-1]^{-1}$ is the only distribution
that is stationary ($[dN_Q/dt]_{\rm scatt}=0$) under collisions.  For weak driving, the scattering
term can be linearized to the form
\begin{eqnarray}
\left(\frac{dN_Q}{dt}\right)_{\rm scatt} &=&-\sum_{Q^\prime} S_{Q,Q^\prime}\Phi_{Q^\prime} \nonumber \\
\Phi_{Q^\prime}(x)&=&N_{Q^\prime}(x)-n_{Q^\prime}(T(x)),
\label{eq:linscat}
\end{eqnarray}
where the linearized scattering operator $S_{Q,Q^\prime}$ is non-negative \cite{Ziman1960}.  
It can be made real-symmetric by multiplying by $n_{Q^\prime}(n_{Q^\prime}+1)$.  Its eigenvalues
are all greater than 0 except for one zero eigenvalue related to conservation of phonon energy.
The rate of change of energy density caused by collisions is
\begin{equation}
0=\left(\frac{dU}{dt}\right)_{\rm scatt}=-\frac{1}{\Omega_S}\sum_{QQ^\prime} \hbar\omega_Q S_{QQ^\prime}\Phi_{Q^\prime}. 
\label{eq:Escatt}
\end{equation}
Since this must hold for any possible deviation $\Phi_Q$ from equilibrium, then $\hbar\omega_Q$ must
be a null left eigenvector of the scattering matrix,
\begin{equation}
\sum_Q \hbar\omega_Q S_{QQ^\prime}=0.
\label{eq:nullEV}
\end{equation}
This relation (which will be invoked later) and other aspects are discussed elsewhere \cite{Allen2018}.  
The diagonal elements $S_{Q,Q}$ are the ``single-mode relaxation rates'' $1/\tau_Q$.

The ``drift'' term in Eq.(\ref{eq:pbe}) has the form
\begin{equation}
\left(\frac{dN_Q}{dt}\right)_{\rm drift}=-\vec{v}_Q \cdot \vec{\nabla}N_Q=-\vec{v}_Q\cdot\left[\frac{dn_Q}{dT}\vec{\nabla}T  
+\vec{\nabla}\Phi_Q \right]
\label{eq:}
\end{equation}
The rate of energy change caused by drift is
\begin{eqnarray}
\left(\frac{dU}{dt}\right)_{\rm drift}&=&\frac{1}{\Omega_S}\sum_Q \hbar\omega_Q \left(\frac{dN_Q}{dt}\right)_{\rm drift} \nonumber \\
&=&-\frac{1}{\Omega_S}\sum_Q \hbar\omega_Q\vec{v}_Q \cdot\vec{\nabla}\Phi_Q = -\vec{\nabla}\cdot\vec{j}.
\label{eq:Udr}
\end{eqnarray}

The external driving term deserves discussion:
\begin{equation}
\left( \frac{dN_Q}{dt}\right)_{\rm ext}=\frac{P(x)}{C} \frac{dn_Q}{dT}.
\label{eq:ext}
\end{equation}
The need for such a term was only recently recognized \cite{HuaB2014,Hua2015,VermeerschI2015},
and was incorrectly omitted in an earlier paper \cite{Allen2014}.  The specific form on the right-hand side of Eq.
\ref{eq:ext} is not unique, but depends on the geometry being modeled.  This version, used in ref.
\onlinecite{VermeerschI2015}, is appropriate for the Zhou {\it et al.} simulation \cite{Zhou2009}: the thermostat is designed
to increase occupancies $N_Q$ of modes at the same rate that a uniform rate of temperature increase $\dot{T}=P/C$
would cause an equilibrated system to increase $n_Q(T(t))$. 
$P(x)$ is the volume rate of heating at spatial point $x$, and $C$ is the volumetric heat capacity.
Then the total volumetric energy input is
\begin{equation}
\left(\frac{dU}{dt}\right)_{\rm ext}=\frac{1}{\Omega_S}\sum_Q \hbar\omega_Q \left(\frac{dN_Q}{dt}\right)_{\rm ext}=P(x),
\label{eq:Uext}
\end{equation}
Energy conservation as given by Eqs. \ref{eq:pbe}, \ref{eq:Escatt}, \ref{eq:Udr}, and \ref{eq:Uext} is  
\begin{equation}
\frac{\partial U}{\partial t}=0=P-\vec{\nabla}\cdot\vec{j}.
\label{eq:Ucons}
\end{equation}
Since we consider steady state situations with time-independent driving, $\partial U/\partial t=0$.

The full linearized PBE now takes the form
\begin{equation}
\vec{v}_Q \cdot \left[ \frac{dn_Q}{dT}\vec{\nabla}T+\vec{\nabla}\Phi_Q\right]
+\sum_{Q^\prime} S_{Q,Q^\prime} \Phi_{Q^\prime} = \frac{P}{C} \frac{dn_Q}{dT}.
\label{eq:fPBE}
\end{equation}
Here the fields $T=T_0+\Delta T(\vec{r})$, $\Phi_Q$, and $P$ are all in coordinate ($\vec{r}$) space.  The equation
simplifies in reciprocal space.  For the one-dimensional version, this is
\begin{eqnarray}
ikv_{Qx} \left[ \frac{dn_Q}{dT} \Delta T(k)+\Phi_Q(k)\right]
&+&\sum_{Q^\prime} S_{Q,Q^\prime} \Phi_{Q^\prime}(k) \nonumber \\
&=& \frac{P(k)}{C} \frac{dn_Q}{dT},
\label{eq:kPBE}
\end{eqnarray}
where $\nabla_x T(x)\rightarrow ik\Delta T(k)$.  
Solution requires inversion of a non-Hermitean $Q$-space matrix $\hat{S}+ikv_x \hat{1}$.
For many purposes it is sufficient to make the ``single-mode relaxation time
approximation'' (RTA), $S_{Q,Q^\prime}\rightarrow \delta_{Q,Q^\prime}/\tau_Q$.
This permits a simple solution,
\begin{equation}
\Phi_Q(k)=-\frac{dn_Q}{dT}\frac{[ikv_{Qx}\Delta T(k)-P(k)/C]}{1/\tau_Q + ikv_{Qx}}.
\label{eq:s1}
\end{equation}
We now want to eliminate the field $P(k)$.  This can be done \cite{Allen2018} using local energy conservation,
as advocated in Refs. \onlinecite{HuaB2014,Hua2015,VermeerschI2015}.  The form used in these papers is
\begin{equation}
\sum_Q \hbar\omega_Q (N_Q-n_Q)/\tau_Q=0,
\label{eq:CC}
\end{equation}
which is the RTA version of Eq. \ref{eq:Escatt}.  This equation is {\bf not} satisfied in RTA for
arbitrary $\Phi_Q$, but it is sensible to require the chosen steady state distribution to satisfy it.
Equivalently, one can use $P(k)=i\vec{k}\cdot\vec{j}(k)$.

From Eqs. \ref{eq:JJ} and \ref{eq:linscat}, the current can be written as
\begin{equation}
j_x(k)=\frac{1}{\Omega_S}\sum_Q \hbar\omega_Q v_{Qx} \Phi_Q(k)
\label{eq:Jk}
\end{equation}
since the equilibrium distribution $n_Q$ carries no current.
Then the current satisfies
\begin{equation}
Z(k)j_x(k)=-\kappa_1(k) \nabla_x T(k)
\label{eq:ZJ}
\end{equation}
where $\kappa_1$ comes from the first term on the right of Eq. \ref{eq:s1},
\begin{equation}
\kappa_1(k)=\frac{1}{\Omega_S}\sum_Q \frac{\hbar\omega_Q v_{Qx}^2 (dn_Q/dT)}{1/\tau_Q+ikv_{Qx}}.
\label{eq:kZ}
\end{equation}
The function $\kappa_1$ is the nonlocal thermal conductivity that comes from incorrectly omitting
the external driving, Eq. \ref{eq:ext}.   It is the phonon analog of the Reuter-Sondheimer theory \cite{Reuter1948}
of the anomalous skin effect contained in the electrical conductivity $\sigma(\vec{k},\omega)$
in the dc ($\omega=0$) limit.

The subscript $1$ in Eq. \ref{eq:kZ} indicates omission
of a  ``renormalization'' factor $1/Z(k)$ (or equivalently, setting $Z$ to 1).  $Z(k)$
contains the effects of the driving term, and has the form 
\begin{equation}
Z(k)=1-\sum_Q \frac{C_Q}{C} \frac{ikv_{Qx}}{1/\tau_Q + ikv_{Qx}},
\label{eq:Z1}
\end{equation}
and $C_Q$ is $\hbar\omega_Q (dn_Q/dT)/\Omega_S$, the contribution to the heat
capacity $C$ from mode Q.  Since $\sum_Q C_Q /C=1$, the renormalization factor
can be written as
\begin{equation}
Z(k)=\sum_Q \frac{C_Q}{C} \frac{1/\tau_Q}{1/\tau_Q + ikv_{Qx}}
=\sum_Q \frac{C_Q}{C} \frac{1}{1 + k^2\ell_{Qx}^2}.
\label{eq:Z}
\end{equation}
where $\ell_{Qx}=v_{Qx}\tau_Q$.  Then
the thermal conductivity in PBE theory with RTA is
\begin{equation}
\kappa(k)=\kappa_1(k)/Z(k).
\label{eq:k}
\end{equation}
This equation is different in appearance but is equivalent to those in Refs. \onlinecite{HuaB2014,VermeerschI2015}.

\section{Debye $\kappa(k)$}
\label{sec:Debye}

In the bulk limit, if $T$ is not too low, the RTA is known to reproduce quite well the true solution, 
if $1/\tau_Q$ is the actual complicated single mode phonon relaxation rate.  For qualitative understanding,
simpler models are desirable.  In the Debye model there are
three acoustic phonon branches, all having the form $\omega_Q=v|\vec{q}|$, all
with the same velocity $v$.  As a supplement to the Debye model, take the relaxation rates $1/\tau_Q$
to have simple power laws in phonon wavevector $|\vec{q}|$,
$(1/\tau_D)(q/q_D)^p$.  Here $q_D$ is the Debye wavevector, and $1/\tau_D$ is a maximum scattering
rate, which depends on $T$, being linear in $T$ at higher $T$.
The scale of $\kappa(T)$ in the Debye model is
\begin{equation}
\kappa_0=\frac{k_B v^2 \tau_D}{\Omega_{\rm cell}},
\label{eq:k0}
\end{equation}
which depends on $T$ because of $\tau_D$.

In Debye approximation, Eqs.(\ref{eq:kZ},\ref{eq:Z}) become, in the classical ($k_B T>\hbar\omega_D$) limit,
\begin{equation}
\kappa_{1D}(k) = \frac{9k_B v^2}{2\Omega_{\rm cell}} \int_0^{Q_D} \frac{dQ Q^2}{Q_D^3}
\int_{-1}^1 \frac{\mu^2 d\mu}{1/\tau_Q+ikv\mu}
\label{eq:kZD}
\end{equation}
\begin{equation}
Z _{D}(k) = \frac{3}{2} \int_0^{Q_D} \frac{dQ Q^2}{Q_D^3}
\int_{-1}^1d\mu  \frac{1/\tau_Q}{1/\tau_Q+ikv\mu}
\label{eq:ZD}
\end{equation}
where $\mu$ is $\cos\theta$, and $\theta$ is the angle between the velocity (parallel to $\vec{q}$)
and the direction of the temperature gradient (parallel to $\vec{k}=k\hat{x}$).  A factor of 3 appears
in Eq.(\ref{eq:kZD}), to account for the three acoustic branches.

There is no complete consensus about what the power $p$ should be.
Herring \cite{Herring1954} advocated $p=2$, and has received experimental confirmation
\cite{Damen1999}.  However, subsequent studies
\cite{Esfarjani2011,Ma2014,Zhou2016} differ somewhat.  Often, 
for ``N'' (Normal, $\sum\vec{q}$ conserved) scattering, $p=2$, while for ``U'' (Umklapp, 
$\sum\vec{q}$ altered by a reciprocal lattice vector $\vec{G}$),  $p=3$.  
For general $p$, Eqs.(\ref{eq:kZD},\ref{eq:ZD}) become
\begin{equation}
\kappa_{1Dp}(k)=\frac{9\kappa_0}{2} \int_0^1 dx \int_{-1}^1 d\mu \frac{x^2 \mu^2}{x^p+iy\mu}
\label{eq:kZp}
\end{equation}
\begin{equation}
Z_{Dp}(k)=\frac{3}{2} \int_0^1 dx \int_{-1}^1 d\mu \frac{x^2 x^p}{x^p+iy\mu}
\label{eq:Zp}
\end{equation}
where $x=q/q_D$ and $y=kv\tau_D=k\ell_{\rm min}$.
Algebraic formulas for these integrals are given in Appendix B.  When $T\gg \Theta_D$ is not
obeyed, Eqs.(\ref{eq:kZp},\ref{eq:Zp}) each need a quantum factor $(x\gamma_T/\sinh(x\gamma_T))$
inside the $x$-integral, where $\gamma_T=\hbar\omega_D/2k_B T$.

\begin{figure}
\includegraphics[angle=0,width=0.5\textwidth]{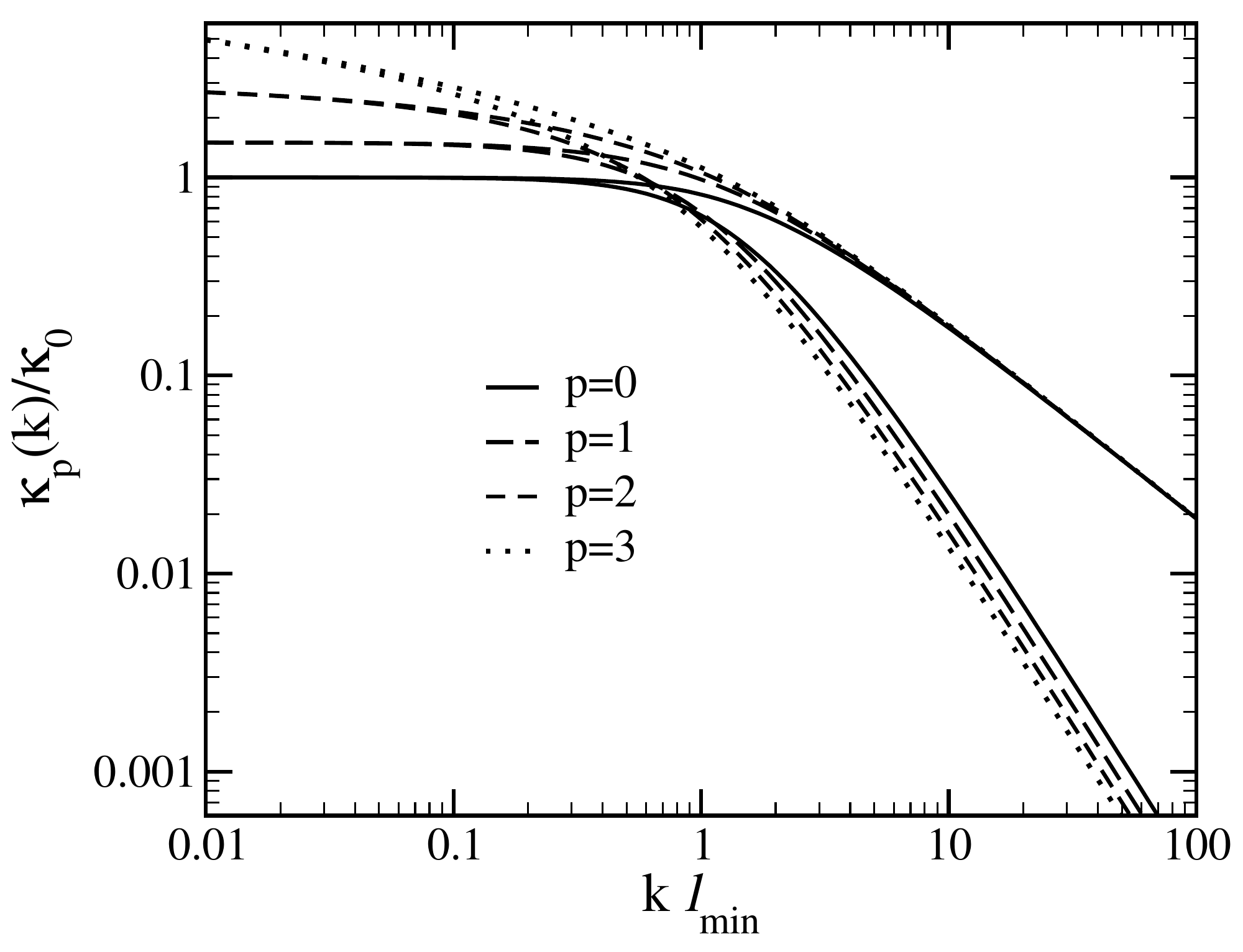}
\caption{\label{fig:kp} Power law Debye models for $\kappa(k)/\kappa_0$ at high $T$.
These are derived from the models $1/\tau_Q\propto Q^p \propto \omega^p$ and $p=0,1,2,3$.  
The four cases each have one curve at small $k\ell_{\rm min}$, which splits into two curves
at large $k\ell_{\rm min}$.  
The upper branches are numerical solutions of the full high $T$
Boltzmann theory in the RTA/Debye model, Eq. \ref{eq:k}.  
The lower branches are $\kappa_1(k)$, Eq. \ref{eq:kZ}, for the same models, 
omitting the renormalization $Z$.
The $p=2$ and $3$ full versions are shown on a linear scale in Fig. \ref{fig:calfit}. }
\end{figure}

The answers simplify at large wavevector $k\ell_{\rm min}>>1$,

\begin{eqnarray}
\kappa_{1Dp}(k)&\rightarrow& \frac{9\kappa_0}{(3+p)(k\ell_{\rm min})^2} \nonumber \\
Z_{Dp}&\rightarrow&\frac{3\pi}{(3+p)2k\ell_{\rm min}} \nonumber \\
\kappa_{Dp}\equiv\kappa_{1DP}/Z_{Dp}&\rightarrow& \frac{6\kappa_0}{\pi k\ell_{\rm min}}
\label{eq:largek}
\end{eqnarray} 
However, Boltzmann theory for the statistical evolution of $N_Q(\vec{r})$ is hard to justify on
atomic distance scales, or at wavevectors $k$ as large as a reciprocal lattice vector $G=2\pi/a$.
\footnote{In metals, a corresponding
Boltzmann equation gives a good theory for susceptibility $\chi(k,\omega)$ at small $k$
and $\omega$, but does not contain Friedel oscillations at $k=2k_F$ or high frequency
plasma oscillations or interband effects.  For phonons, there is no analog of either Friedel or plasma oscillations,
so breakdown at large $k,\omega$ is probably gradual.} 
The secure small $k\ell_{\rm min}=y$ part of the formulas for $\kappa_{1Dp}$ (Eq. \ref{eq:kZp}) is
\begin{eqnarray}
&&(p=0) \ \ \kappa_{1D0}(k)\sim\kappa_0(1-3y^2/5+\ldots) \nonumber \\
&&(p=1) \ \ \kappa_{1D1}(k)\sim\frac{3}{2}\kappa_0\left(1-\frac{6}{5}y^2 \log\frac{1}{y}+\ldots\right) \nonumber \\
&&(p=2) \ \ \kappa_{1D2}(k)\sim 3\kappa_0 \left(1-\frac{3\pi}{7}\sqrt\frac{y}{2}+\ldots \right) \nonumber \\
&&(p=3) \ \ \kappa_{1D3}(k)\sim \kappa_0 \left(\log\frac{1}{y} + \ldots\right) 
\label{eq:ksmq}
\end{eqnarray}
Notice that the small $k$ parts for $p\ge 1$ have non-analytic $k$-dependences, and the
$p=3$ formula diverges logarithmically.  Similarly, the small $k$ results for $Z_{Dp}$ (Eq. \ref{eq:Zp}) are
\begin{eqnarray}
&&(p=0) \ \ Z_{D0}(k)\sim 1-y^2/3+\ldots) \nonumber \\
&&(p=1) \ \ Z_{D1}(k)\sim 1-y^2 + \ldots \nonumber \\
&&(p=2) \ \ Z_{D2}(k)\sim 1-\frac{3\pi\sqrt{2}}{10} y^{3/2} +\ldots \nonumber \\
&&(p=3) \ \ Z_{D3}(k)\sim  1-\pi y/4 + \ldots 
\label{eq:Zsmq}
\end{eqnarray}
Full results from Eqs.(\ref{eq:kZp},\ref{eq:Zp}) are in Fig. \ref{fig:kp}.  The full theories
$\kappa_{Dp}=\kappa_{1Dp}/Z_{Dp}$ agree with the unrenormalized results $\kappa_{1Dp}$ at small
$k$.  Notice that at large $k$, in agreement with Eq. \ref{eq:largek}, the full theories converge to an 
answer independent of exponent $p$.
This relates to the fact that large $k$ corresponds to small distances where results should not depend on 
mean free paths, which are all longer than the distance scale.  However, the unrenormalized curves
fall off faster with $k$ and shift depending on $p$.
\par
\par

\section{Callaway $\kappa(k)$}
\label{sec:kC}

Callaway \cite{Callaway1959,Allen2013} devised an improved version of the relaxation time approximation.
Because of anharmonic terms in the interatomic potential,
a phonon $Q$ with wavevector $\vec{q}$ can decay into two phonons of wavevector $\vec{q}_1+\vec{q}_2$
provided $\vec{q}=\vec{q}_1+\vec{q}_2+\vec{G}$.  The N processes have the reciprocal lattice vector 
$\vec{G}=0$, and the U processes have $\vec{G}\ne 0$. Peierls \cite{Peierls1929} pointed
out that N processes cannot fully relax the heat current.
This is particularly important at lower $T$, because U processes require
higher energy phonons and are thus suppressed at lower $T$.  

Callaway's model for the rate of change of the phonon distribution $N_Q$ is
\begin{equation}
\left(\frac{dN_Q}{dt}\right)_{\rm collision}=-\frac{N_Q-n_Q}{\tau_{QU}}-\frac{N_Q-n_Q^\ast}{\tau_{QN}}
\label{eq:Ca}
\end{equation}
The distribution $n_Q^\ast$ is the one which maximizes entropy subject to conservation of
both energy and wavevector.  It is a modified Bose-Einstein distribution
\begin{equation}
n_Q^\ast=\frac{1}{\exp(\hbar\omega_Q/k_B T +\Lambda_x q_x)-1},
\label{eq:nstar}
\end{equation}
where $\Lambda_x$ is a Lagrange multiplier fixed by the condition \footnote{Callaway 
used a slightly different and less correct condition; see Ref. \onlinecite{Allen2013}}
 $\sum_Q q_x n_Q^\ast=q_{x,\rm tot}=\sum_Q q_x \Phi_Q$.
The natural extension of Debye-type relaxation laws in the Callaway scheme is
 \begin{eqnarray}
1/\tau_{Q}&=&1/\tau_{QU}+1/\tau_{QN}=(1/\tau_D)(r_U x^3 + r_N x^2) \nonumber \\
1/\tau_{QU}&=&1/\tau_U (q/q_D)^3 = (1/\tau_D) r_U x^3 \nonumber \\
1/\tau_{QN}&=&1/\tau_N (q/q_D)^2 = (1/\tau_D) r_N x^2
\label{eq:cscat}
\end{eqnarray}
where $r_U$ and $r_N$ are the relative rates of U and N scattering, with $r_U+r_N=1$.
The coefficients $r_U$ and $r_N$ depend strongly on $T$ at low $T$, but are $T$-independent at
higher $T$ (where $1/\tau_D \propto T$).

Zhou {\it et al.} \cite{Zhou2016} computed relaxation rates for GaAs
in the classical limit.  The results in Fig. 2 of their paper
indicate that $r_U\sim 0.9$ and $r_N\sim 0.1$.  One could expect similar values for GaN.
The Boltzmann equation (\ref{eq:kPBE}) can be written in Callaway form and solved.
The Callaway result (with both $r_N$ and $r_U$ non-zero)
cures the logarithmic divergence ($\kappa\propto\log L$) obtained when purely $p=3$
Umklapp scattering is used, and gives a finite $\kappa(k\rightarrow 0)$ limit.
Appendix C contains the derivation of the Callaway correction to the nonlocal theory for $\kappa(k)$.
Unfortunately, the data shown in Fig. \ref{fig:calfit} do not extend to low enough $k$ to 
enable a choice to be made about actual relaxation rates and how they are distributed 
between N and U processes.

\section{extrapolating $\kappa$ to $L\rightarrow\infty$}
\label{sec:extrap}

\begin{figure}
\includegraphics[angle=0,width=0.5\textwidth]{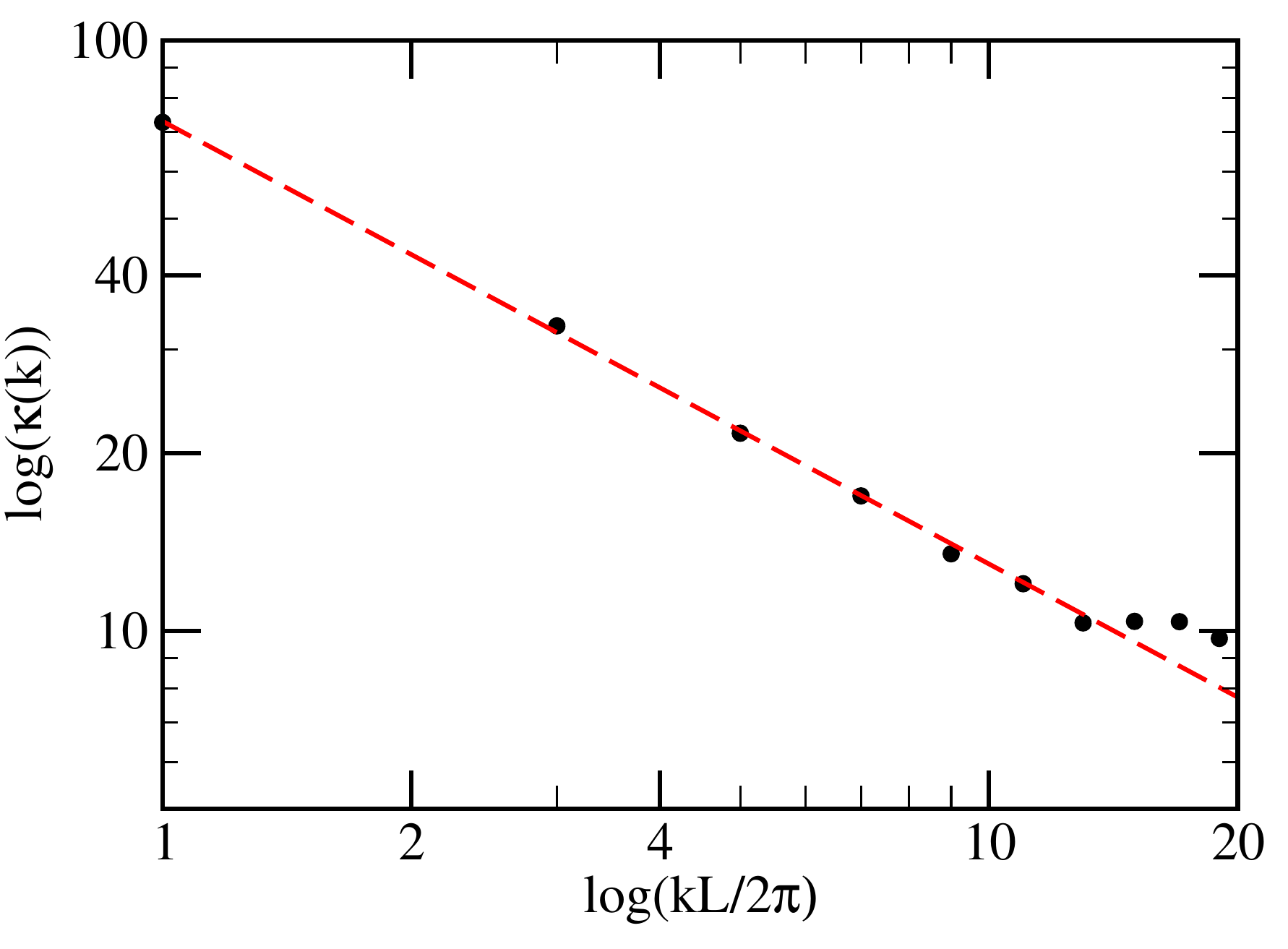}
\caption{\label{fig:logkap} The $\kappa(k)$ data from Fig. \ref{fig:calfit}
are plotted here on a logarithmic scale, showing an approximate $\kappa\propto 1/k^{0.75}$
fit.  This diverges strongly as $k\rightarrow 0$.  The accuracy of this fit is probably accidental, because
there is no theory to justify it. }
\end{figure}

Zhou {\it et al.} \cite{Zhou2009} attempt extrapolation of their finite size ($L$) simulations
to $L\rightarrow\infty$, and notice difficulties.  The present results
require alternate extrapolations.  One way is to choose a model relaxation
time and use the resulting Debye RTA theory to fit the $\kappa(k)$ curves in Fig. \ref{fig:calfit}.
Three things should be stressed. First, the theoretical curves do not give a
particularly good fit to the higher $k$ part of $\kappa(k)$.
This is not surprising.  Debye approximation describes small $Q$ phonon properties,
but does not recognize the small group velocities and corresponding small mean free paths
of optical modes.  These modes do carry heat, and are not suppressed at larger $k$. 
Second, a surprisingly nice fit with no theory at
all can be made by plotting $\log \kappa(k)$ {\it versus} $\log k$.  This is shown in
Fig. \ref{fig:logkap}.
The result, that $\kappa(k)$ is roughly proportional to $k^{-0.75}$, should not be taken
seriously, even though it is as good a fit as any obtained from Debye RTA theory.  A $k^{-0.75}$
divergence implies a scaling $\kappa_{\rm bulk} \propto L^{0.75}$ which has never been
detected experimentally, and is almost certainly unphysical.
Nevertheless, it is an intriguing observation which might motivate further
investigation of behavior of $\kappa(k)$ in vibrational heat conductors.  The third thing
to be stressed is that fitting the $\kappa(k)$ data is an imperfect enterprise.  The numerical $T(x)$ data
of Zhou {\it et al.} shown in Fig. \ref{fig:Tofx} were not originally intended for this purpose.
The $k$-points for which numbers can be found are too sparse for confident fits, and are
affected by noise of computation and digitization.  

\par
\par

Nevertheless, theory makes relevant points.  The fits shown in Fig. \ref{fig:calfit}
provide understanding of the unexpectedly fast increase of $\kappa(k)$ as $k$ decreases.
They also provide two kinds of guidance for extrapolation.  First there is direct use of theory
and numerics for $\kappa(k)$.  The two curves in Fig. \ref{fig:calfit}, when extended to $k=0$, yield
\begin{eqnarray}
&&(p=2) \ \ \kappa(0)=3\kappa_0\sim 240{\rm \ W/mK} \nonumber \\
&&(p=3) \ \ \kappa(0)=\infty\times\kappa_0 \sim\infty \nonumber \\
\label{eq:kn0}
\end{eqnarray}
Evidently extrapolation to $L\rightarrow \infty$ is even more uncertain than 
imagined by Zhou {\it et al.}.  

The other version of extrapolation indicated by this analysis is, following Zhou {\it et al.},
to plot $\kappa(L)$ obtained from the temperature slope at mid-point, against 
various functions of $L$.  Zhou {\it et al.} used $\kappa(\infty)-\kappa^\prime /L$.  Formulas derived here show
that $\kappa(\infty)-\kappa^\prime/\sqrt{L}$ has more theoretical justification and better correspondence with
the computed $\kappa(k)$.

\par
\begin{figure}
\includegraphics[angle=0,width=0.5\textwidth]{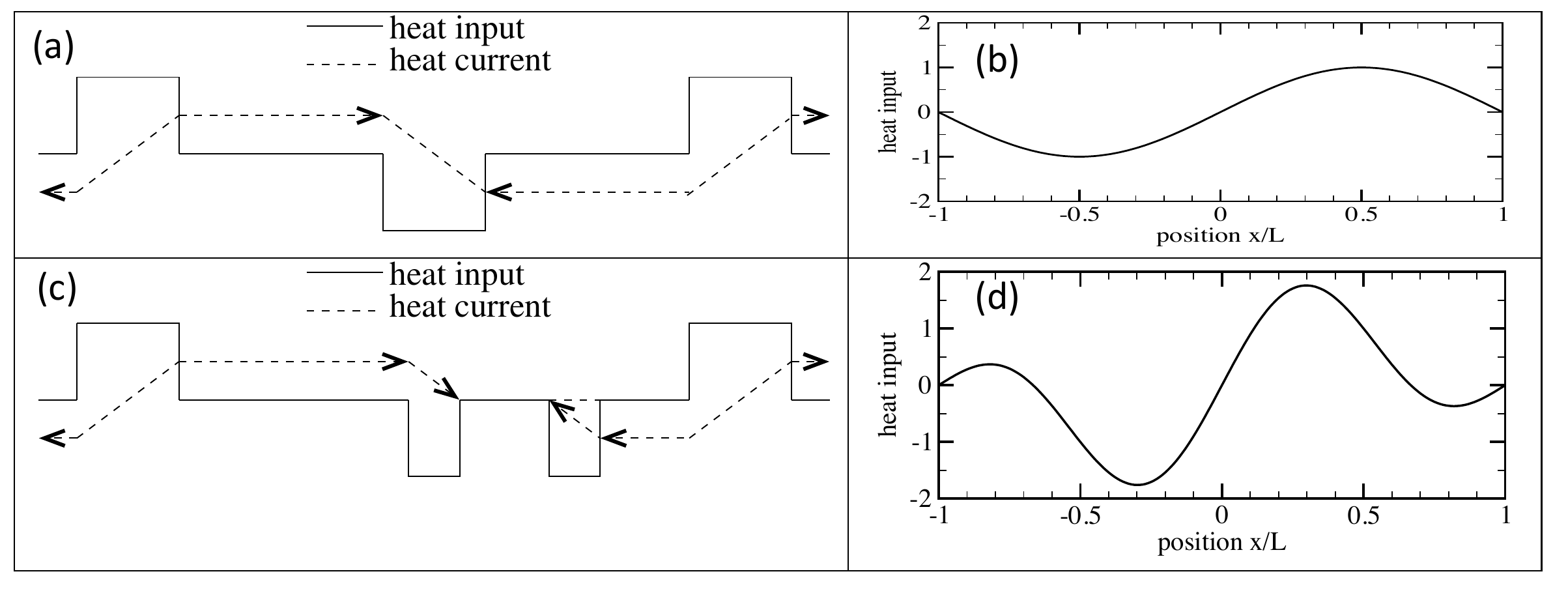}
\caption{\label{fig:heatsines} Choice (a) is standard and used in Fig. \ref{fig:Tofx}.
The others are suggested alternative heating profiles  The sine curve (b) is the simplest.
The asymmetrical block heating (c) allows all Fourier components to be extracted.  
Two sines (curve (d)) allows $k=2\pi/L$ and $4\pi/L$ to be extracted simultaneously. }
\end{figure}
\par

Finally, Fig. \ref{fig:heatsines} suggests alternate ways of performing NEMD simulations.
A more rapid reduction of noise will happen if instead of insertion of heat into separate
isolated regions (panel a), the heat is inserted sinusoidally (panel b).  
This has been tested \cite{Yerong2016} with some success
for a simple model.  But this gives only a single $k$-point for $\kappa(k)$, while a mesh of small-$k$
points contains much additional insight.  The heating pattern could use more than one
sinusoidal period, as in Fig. \ref{fig:heatsines}d.  Finally, it is frustrating that the analysis
done here yields only $k_n$ with odd integer $n$; if $n=2,4,6$ were available to
supplement $n=1,3,5$, then theoretical fits could be judged with more confidence.  The
even integers were excluded by the mirror symmetry of the heat input.  For example, the mirror
in panel (a) is around $x=0$, and in panel (b), around $x=L/4$.
Arrangements like those shown in panels (c) and (d) provide the desired symmetry breaking.

\section{acknowledgements}
I am grateful to Chengyun Hua, Mengkun Liu, A. J. H. McGaughey, A. J. Minnich, V. Perebeinos,
P. K. Schelling, and Xiaowang Zhou for useful advice.  
I thank M. V. Fernandez-Serra and J. Siebert for inspiration.
This work was supported in part by DOE grant No. DE-FG02-08ER46550.

\section{appendix A: Boundary effects in Simulations}
The characteristic sigmoid shape seen in the Zhou {\it et al.} \cite{Zhou2009} simulations of
Fig. \ref{fig:Tofx} is not always as prominent in other simulations.  Comparison of various simulations
indicates that the detailed expression of non-locality differs depending on boundary conditions.
For slab problems, either ``transparent'' or ``opaque'' boundary conditions are used.  With transparent boundaries, 
the unperturbed simulation cell is repeated periodically in all three directions.  Heat 
is added and removed somewhere in the cell interior.   
With opaque boundaries, periodicity of the simulation cell in the direction of heat flow
is irrelevant.  Homogeneity is broken, and atoms near the boundary have to respond, not to a 
periodic image, but to the actual heated boundary.  Analysis of such situations \cite{Hua2015}
(which can be experiment or simulation) requires a model of how the boundary emits and
reflects vibrations.  The temperature distributions near the boundary can therefore vary.
This paper deals only with the simpler transparent version.   Other examples (besides Zhou {\it et al.}) 
of simulations with transparent boundaries are Aubry {\it et al.} \cite{Aubry2008},
Goel {\it et al.} \cite{Goel2015}, and Gordiz and Henry \cite{Gordiz2017}.  
Examples of opaque boundaries are Landry and McGaughey \cite{Landry2009},
Jiang {\it et al.} \cite{Jiang2009}, Cao and Qu
\cite{Cao2012}, and Feng {\it et al.} \cite{Feng2017}.

%
\section{Appendix B: Analytic Integrations}

For integer $p$, the integrations in Eqs.(\ref{eq:kZp},\ref{eq:Zp}) can be done analytically:
\begin{equation}
\frac{\kappa_{1D0}(k)}{\kappa_0}=\frac{3}{y^2}\left[1-\frac{\tan^{-1}(y)}{y} \right]
\label{eq:k0}
\end{equation}
\begin{equation}
\frac{\kappa_{1D1}(k)}{\kappa_0}=\frac{9}{10}\left[ 1+\frac{2}{y^2}
\left(1-\frac{\tan^{-1}(y)}{y}\right)
-y^2 \log\left(1+\frac{1}{y^2}\right)\right]
\label{eq:k1}
\end{equation}
\begin{eqnarray}
\frac{\kappa_{1D2}(k)}{\kappa_0}&=&\frac{9}{7y^2}\left\{ 1-\frac{\tan^{-1}(y)}{y}
+2y^2 \right. \nonumber \\
&-&  \frac{y^{5/2}}{\sqrt 2} \left[\tan^{-1}\left(\sqrt{\frac{2}{y}}+1\right)
+\tan^{-1}\left(\sqrt{\frac{2}{y}}-1\right)\right] \nonumber \\
&-&\left. \frac{y^{5/2}}{2\sqrt 2}  \log\left(\frac{1+\sqrt{2y}+y}
{1-\sqrt{2y}+y}\right)\right\}
\label{eq:k2}
\end{eqnarray}
\begin{equation}
\frac{\kappa_{1D3}(k)}{\kappa_0}=\frac{1}{2}\log\left(\frac{1+y^2}{y^2}\right)
+\frac{1}{y^2}\left(1-\frac{\tan^{-1}(y)}{y}\right)
\label{eq:k3}
\end{equation}
\begin{equation}
Z_{D0}=\frac{\tan^{-1}y}{y}
\label{eq:Z0}
\end{equation}
\begin{equation}
Z_{D1}=\frac{1}{4}-\frac{3y^2}{4}+\frac{3\pi y^3}{8}+\frac{3}{4}(1-y^4)\frac{\tan^{-1}y}{y}
\label{eq:Z1}
\end{equation}
\begin{eqnarray}
Z_{D2}&=&\frac{1}{20y} \left\{ 8y+12\tan^{-1}y-6\sqrt{2} y^{5/2}
\left[\tan^{-1}\left(1+\sqrt{\frac{2}{y}}\right) \right. \right. \nonumber  \\
&+& \left. \left. \tan^{-1}\left(1-\sqrt{\frac{2}{y}}\right)\right] 
+3\sqrt{2}y^{5/2}\log\left(\frac{1+\sqrt{2y}+y}{1-\sqrt{2y}+y} \right) 
\right\} \nonumber \\
\label{eq:Z2}
\end{eqnarray}
\begin{equation}
Z_{D3}=\frac{1}{2}-\frac{\pi y}{4}+(1+y^2)\frac{\tan^{-1}y}{2y}
\label{eq:Z3}
\end{equation}
These equations are plotted versus $y=k\ell_{\rm min}$ in Fig. \ref{fig:kp}. 

\section{Appendix C: Nonlocal Callaway Model}

In the full PBE, both ``Normal'' (N) and ``Umklapp'' (U) scattering events are contained
in the linearized scattering operator $S_{Q,Q^\prime}$.  The total crystal momentum
$\sum_Q \vec{q}N_Q$ is automatically conserved under N-scattering alone.  That is,
if only N terms of $S_{Q,Q^\prime}$ are kept, $\sum_Q\vec{q}N_Q$ is conserved.
At low $T$, where U-scattering
is suppressed relative to N, it is important to recognize the different effects of these two
types of events.  Callaway therefore introduced a modified version of the relaxation time
approximation,
\begin{equation}
\left(\frac{ dN_Q}{dt} \right)_{\rm scatt} \rightarrow -\frac{N_Q -n_Q}{\tau_{QU}}
-\frac{N_Q-n_Q^\ast}{\tau_{QN}}
\label{eq:Cscatt}
\end{equation}
where
\begin{equation}
n_Q^\ast = \frac{1}{\exp(\hbar\omega_Q/k_BT+\vec{q}\cdot\vec{\Lambda})-1}
\approx n_Q-\frac{dn_Q}{dT}\frac{k_BT^2}{\hbar\omega_Q}\vec{q}\cdot\vec{\Lambda},
\label{eq:nQN}
\end{equation}
and $\vec{\Lambda}$ is a Lagrange multiplier, adjusted so that the distribution
$n_Q^\ast$ contains all the crystal momentum.  That means
\begin{equation}
\sum_Q \vec{q}N_Q= \sum_Q \vec{q} n_Q^\ast
\label{eq:fixL0}
\end{equation}
or, since the equilibrium distribution $n_Q$ has no net crystal momentum,
\begin{equation}
\sum_Q\vec{q}\Phi_Q = 
-\sum_Q \frac{dn_Q}{dT}\frac{k_BT^2}{\hbar\omega_Q}\vec{q}\vec{q}\cdot\vec{\Lambda}
\label{eq:fixL}
\end{equation}
The total single-mode scattering rate $1/\tau_Q = S_{QQ}$ is
\begin{equation}
1/\tau_Q = 1/\tau_{QU} + 1/\tau_{QN}.
\label{eq:tautot}
\end{equation}
Because of the last term of Eq. \ref{eq:nQN}, the Boltzmann equation has an 
additional term, and the RTA solution, Eq. \ref{eq:s1} takes the form
\begin{equation}
\Phi_Q(k)=-\frac{dn_Q}{dT}\frac{[\vec{v}_{Q}\cdot\vec{\nabla} T(k)+\frac{k_B T^2}{\hbar\omega_Q\tau_{QN}}
\vec{q}\cdot\vec{\Lambda}-\frac{P(k)}{C}]}{1/\tau_Q + i\vec{k}\cdot\vec{v}_{Q}}.
\label{eq:s1new}
\end{equation}

Now it is necessary to have two additional equations, because the two extra fields $P(k)$
and $\vec{\Lambda}(k)$ need to be eliminated.  These equations are energy conservation
(Eqs. \ref{eq:CC} and \ref{eq:Jk}) for eliminating $P$ and crystal momentum conservation
(Eq. \ref{eq:fixL}) for $\vec{\Lambda}$.  The result of using
Eq. \ref{eq:CC} to eliminate $P/C$ is a minor extension of the previous result, Eq. \ref{eq:k},
containing the same renormalization factor $Z$, in Eq. \ref{eq:Z}.  To avoid $3\times 3$ matrix
equations, orthorhombic or higher symmetry is now assumed, and one-dimensional transport
along an orthorhombic axis denoted $x$.  This permits $\vec{q}\cdot\vec{\Lambda}$
to be simplified to $q_x \Lambda$.  The current density is
\begin{equation}
j_x = -\frac{\kappa_1}{Z} \frac{dT}{dx}-\frac{k_B T^2}{\Omega_S Z}\sum_Q \frac{dn_Q}{dT}
\frac{v_{Qx}/\tau_{QN}}{1/\tau_Q +ikv_{Qx}}q_x \Lambda.
\label{eq:JLam}
\end{equation}
 Here $\kappa_1$ is the same as before, Eq. \ref{eq:kZ}, with both N and U processes included in
 the scattering $1/\tau_Q$ as in Eq. \ref{eq:tautot}.  The second term in Eq. \ref{eq:JLam} is
 an additional current that comes from the fact that the scattering caused by N processes
 has been overestimated in the first term.  The new formula for the distribution
 function is
 \begin{equation}
\Phi_Q=-\frac{1}{Z}\frac{dn_Q}{dT}\frac{v_{Qx}\nabla T
+\frac{k_B T^2}{\hbar\omega_Q\tau_{QN}}q_x\Lambda}{1/\tau_Q+ikv_{Qx}}.
\label{eq:PhiL}
\end{equation}

Now use Eq. \ref{eq:fixL} to eliminate the Lagrange multiplier $\Lambda$. After some algebra,
the result for the Callaway heat conductivity $\kappa_C(k)$ can be written
\begin{equation}
\kappa_C(k)=\frac{\kappa_1}{Z}+\frac{1}{Z\Omega_S}
\frac{\sum_Q H_{Qx}\frac{q_x}{\tau_{QN}} \sum_{Q^\prime} H_{Q^\prime x}q_x^\prime}
{\sum_Q\frac{dn_Q}{dT} \frac{q_x^2}{\hbar\omega_Q}\left[Z-\frac{1/\tau_{QN}}{1/\tau_Q +ikv_{Qx}}\right]},
\label{eq:kC}
\end{equation}
where
\begin{equation}
H_{Qx}=\frac{dn_Q}{dT} \frac{v_{Qx}}{1/\tau_Q +ikv_{Qx}}.
\label{eq:HQ}
\end{equation}
The second term of Eq. \ref{eq:kC} is the Callaway correction to the nonlocal RTA thermal conductivity
(when the thermal variation is one-dimensional, and the symmetry orthorhombic or higher.)  In the local limit
$k\rightarrow 0$ and $Z\rightarrow 1$, this answer agrees exactly with Eq. 15 of Ref. \onlinecite{Allen2013}.  


\bibliography{2018GaNpaper}

\end{document}